\documentclass[12pt]{article}
\usepackage{amsmath, amssymb, amsfonts, amsthm}
\usepackage{graphicx, placeins}
\usepackage{enumerate, enumitem}
\usepackage{natbib} 
\usepackage{url, hyperref} 
\usepackage{algorithm, algorithmic}
\usepackage[font=small,skip=0pt]{caption}
\usepackage{booktabs, multirow}
\usepackage{float}
\usepackage{array}
\usepackage{pdflscape}
\usepackage{xcolor}
\newcommand{\blind}{0}

\usepackage{setspace}
\hypersetup{
  colorlinks   = true, 
  urlcolor     = black, 
  linkcolor    = black, 
  citecolor   =  black 
}

\begin{document}

\def\spacingset#1{\renewcommand{\baselinestretch}%
{#1}\small\normalsize} \spacingset{1}


\if0\blind
{
  \title{\bf Evaluation of adaptive sampling methods in scenario generation for virtual safety impact assessment of pre-crash safety systems}
  \author{Xiaomi Yang\thanks{Corresponding author: Xiaomi Yang, email: xiaomi.yang@chalmers.se}\\
    Division of Vehicle Safety, Chalmers University of Technology \\
    and \\
    Henrik Imberg \\
    Department of Mathematical Sciences, \\
    Chalmers University of Technology
    and University of Gothenburg \\
    and \\
    Carol Flannagan \\
    University of Michigan Transportation Research Institute \\
    Division of Vehicle Safety, Chalmers University of Technology \\
    and \\
    Jonas Bärgman \\
    Division of Vehicle Safety, Chalmers University of Technology}
     \date{} 
  \maketitle
} \fi

\if1\blind
{
  \bigskip
  \bigskip
  \bigskip
  \begin{center}
    {\LARGE\bf Title}
\end{center}
  \medskip
} \fi

\bigskip
\newpage
\begin{abstract}

Virtual safety assessment plays a vital role in evaluating the safety impact of pre-crash safety systems such as advanced driver assistance systems (ADAS) and automated driving systems (ADS). However, as the number of parameters in simulation-based scenario generation increases, the number of crash scenarios to simulate grows exponentially, making complete enumeration computationally infeasible. Efficient sampling methods, such as importance sampling and active sampling, have been proposed to address this challenge. However, a comprehensive evaluation of how domain knowledge, stratification, and batch sampling affect their efficiency remains limited. 

This study evaluates the performance of importance sampling and active sampling in scenario generation, incorporating two domain-knowledge-driven features: adaptive sample space reduction (ASSR) and stratification. Additionally, we assess the effects of a third feature, batch sampling, on computational efficiency in terms of both CPU and wall-clock time. Based on our findings, we provide practical recommendations for applying ASSR, stratification, and batch sampling to optimize sampling performance. 

Our results demonstrate that ASSR substantially improves sampling efficiency for both importance sampling and active sampling. When integrated into active sampling, ASSR reduces the root mean squared estimation error (RMSE) of the estimates by up to 90\%. Stratification further improves sampling performance for both methods, regardless of ASSR implementation. When ASSR and/or stratification are applied, importance sampling performs on par with active sampling, whereas when neither feature is used, active sampling is more efficient. Larger batch sizes reduce wall-clock time but increase the number of simulations required to achieve the same estimation accuracy. 

In conclusion, applying ASSR and stratification in importance sampling and active sampling, where applicable, significantly improves efficiency, enabling the reallocation of computational resources to other safety initiatives. 

\end{abstract}

\noindent%
\vfill

\newpage
\spacingset{1.3} 
\section{Introduction}
\label{sec:intro}
Approximately 1.2 million people die in traffic annually \citep{road_safety_2023}. To address this, both pre-crash safety systems and in-crash protection systems have been developed to prevent or mitigate the consequences of crashes \citep{cicchino2016effectiveness,eichberger2011potentials,jermakian2011crash, evans1986effectiveness, crandall2001mortality}. Verifying the performance of safety systems is essential before their market deployment and throughout the development process. Increasingly, these assessments are performed virtually through simulations \citep{wimmer2023harmonized}.

One virtual assessment method is counterfactual simulations, in which the specific system under assessment is virtually applied to a set of baseline crash kinematics, and the outcomes (e.g., number of crashes, impact speed, or injury risk) are compared between the baseline crashes and the simulations where the pre-crash safety systems are implemented. Consequently, baseline crashes are a prerequisite for counterfactual simulations. More precisely, a set of pre-crash kinematics is required to describe how the involved road users moved prior to the crash---without the safety system under assessment. One source of these kinematics is original reconstructed crashes from in-depth crash databases \citep{rosen2013autonomous,stark2019quantifying,deng2013traffic,yang2024evaluation}. However, these databases are often small due to the cost of in-depth crash investigations. To create a larger set of crashes that better cover more of the possible parameter space and more comprehensively reflect real traffic scenarios\citep{riedmaier2020survey}, new crashes can be generated from those reconstructed crashes. This can be done by applying computational crash-causation-behavior models and critical-event-response-behavior models to them. For example, \citet{bargman2024methodological} validated a combination of crash-causation and critical-event-response models to generate new rear-end crashes. Their crash-causation model is in part based on drivers’ off-road glances and their tendency to brake less aggressively than vehicle's full braking capacity, even in crash situations. This type of counterfactual simulation is called a behavior-model-based counterfactual simulation. 
Behavior-model-based counterfactual simulations typically start with a relatively small number of reconstructed crashes where the evasive maneuver of one of the road users has been removed \citep{rosen2013autonomous}. Each of these events is referred to as a prototype event. The prototype event is then simulated in a virtual environment, with the driver model replacing the removed maneuver. A range of events can be constructed by varying the model parameters, for example: vehicle speed \citep{L3Pilot2021}, driver behavior (e.g., glance behavior and brake deceleration; \citeauthor{lee2018safe}, 2018), and environmental conditions \citep{ruiz2018learning}.

A common problem with scenario generation for counterfactual simulations is that the number of potential crash scenarios increases exponentially as the number of parameters being varied increases. Given the already high computational cost of running the simulations, complete enumeration becomes practically infeasible---and subsampling becomes necessary. Many researchers turn to importance sampling methods to reduce the number of samples while obtaining accurate estimates \citep{wang2021combining,de2017assessment,zhao2016accelerated}. Importance sampling is a variance reduction technique employed in subsampling problems to increase estimation precision \citep{tokdar2010importance}. In the context of scenario generation, the input required for this technique is a proposal distribution on the parameters used, which is a probability distribution that determines the likelihood of sample selection, guiding the sampling process towards more important and informative regions of the input space. A random sample of scenarios is then generated according to the proposal distribution, and an unbiased estimate of the characteristic of interest can subsequently be obtained by inverse probability weighting \citep{mansournia2016inverse}. Ideally, the importance sampling probabilities should be proportional to the outcome of interest for optimal performance \citep{bugallo2017adaptive}. However, specifying the proposal distribution is challenging because the actual outcome is unknown before running the simulation, often resulting in suboptimal performance. Consequently, efficient sampling strategies to improve upon traditional importance sampling methods are needed.

\citet{Imberg2024} introduced active sampling, a machine-learning-assisted subsampling method that combines adaptive importance sampling with predictive modeling to optimize sample selection. While this method addresses many of the limitations inherent in traditional importance sampling, certain aspects and extensions of active sampling in the context of crash-causation scenario generation remain unexplored. For instance, the potential benefits of incorporating domain knowledge to reduce the sample space and minimize the number of simulations required for accurate safety impact assessment have not been investigated. Additionally, previous work by \citet{Imberg2024} focused on the average safety impact over the entire input space without considering stratification, leading to an imbalance among the original prototype events, as different prototype events generate different number of crashes under varying scenarios. 
An important consideration in safety impact assessment is that each prototype event should contribute equally to the evaluation, as these prototypes represent an empirical distribution of crash scenarios and thus should be weighted equally. Case weighting (through post-stratification) and stratification techniques could help ensure a balanced representation of the prototype events in the overall safety impact assessment. Stratification has been shown to reduce variance in stochastic simulations \citep{park2024strata}, yet the relative efficiency of stratification versus post-stratification in crash-causation scenario generation remains unclear. Finally, \citet{Imberg2024} did not explore the potential reduction in overall computational time that could be achieved through parallel computing---an approach that could be particularly beneficial for large-scale virtual safety assessment. To address these gaps, this study evaluates the impact of three previously unexplored features: domain knowledge integration, post-stratification versus stratification, and parallel computing, for both importance sampling and active sampling. Despite its limitations, importance sampling remains widely used and is often easier to implement than active sampling, making it an important benchmark for comparison.

\subsection{Aim}
\label{sec:aim}

The overall aim of this work is to evaluate three implementation features of two adaptive sampling methods. To determine the effectiveness of the features, we applied them to scenario generation-based virtual impact assessments of a pre-crash safety system. Specifically, we aim to
\begin{itemize}
    \item Assess the sampling efficiency of proposed adaptive sample space reduction (ASSR) logic, which applies domain-knowledge-based logical constraints to simulation outcomes, in the context of importance sampling and active sampling. 
    \item Evaluate the effectiveness of stratification within importance sampling and active sampling. 
    \item Analyze the impact of batch size on efficiency in active sampling, particularly in parallel computing environments.
    \item Provide practical guidelines for utilizing ASSR, stratification, and batch sampling in importance sampling and active sampling for scenario generation in virtual safety impact assessments.
\end{itemize}

\section{Methods}
\label{sec:method}
    
Data, models, and simulation setup are described in Section \ref{sec:data}. The implementation of the three features---ASSR logic, stratification, and batch size---is detailed in Section \ref{sec:sampling_methods}. The criteria for sampling stopping conditions are presented in Section \ref{sec:stopping conditions for sampling}. Finally, simulation procedures and performance metrics are described in Section \ref{sec:performance_evaluation}.

\subsection{Data, models, and simulation setup}
\label{sec:data}

The data used for scenario generation in this study consist of the reconstructed pre-crash kinematics of 44 rear-end crashes from a crash database provided by Volvo Car Corporation. The Volvo Cars Traffic Accident Database (VCTAD) contains information about crashes involving Volvo vehicles that occurred in Sweden with repair costs exceeding €4,000 \citep{isaksson2005thirty}. The 44 crashes were selected from a total of 344 rear-end frontal crashes recorded between 2006 and 2017 on roads with a speed limit of $\ge70$ km/h (highways or expressways).

For this work, the braking action by the driver of the following vehicle (FV; the striking vehicle) in each prototype event was removed and replaced by a driver response model. This model triggers braking using a threshold on looming (optically defined time-to-collision, or TTC) as the onset of driver braking \citep{lee1976theory,bargman2024methodological}. The threshold was selected based on work by \citet{markkula2016farewell}. In addition to the driver response model, the counterfactual model for scenario generation in this study included a crash-causation model consisting of two sub-models: an off-road glance-based crash causation sub-model, and a deceleration-based crash causation sub-model. As noted, the latter is based on the fact that, even in crash situations, drivers do not always brake to the full capacity of their vehicles \citep{bargman2024methodological}. See Section \ref{sec:deceleration_crash_causation_model} for details.

The off-road glance-based crash causation sub-model is based on research showing that when the driver’s eyes are on the road ahead, the driver is typically able to react to looming of the lead vehicle (LV), braking early enough and hard enough to avoid a crash. As mentioned, in this crash-causation model-based scenario generation, the main causation factors are drivers’ eyes-off-road (EOFF) glance durations in everyday (non-critical) driving and the braking intensity just prior to crashing. Both factors have been identified as important contributors to the occurrence of rear-end crashes \citep{wang2022causation}. Naturally, glance data are required as input to such a model. Similarly, data describing drivers’ braking behavior, particularly the maximum deceleration reached before impact, are needed to appropriately model driver braking as part of the crash-causation model. In the combined glance and deceleration-based crash-causation model, we treat glances and deceleration as independent variables, as there is no evidence indicating a correlation between the two.

\subsubsection{An EOFF-based crash-causation sub-model}
\label{sec:EOFF_crash_causation_model}

In this study, the EOFF distribution is based on baseline epochs (30s segments) from the Victor et al. (2015) study, which investigated the relationship between the LV driver's visual inattention and distraction (operationalized as off-road glances) and crash risk. EOFF glances were extracted from these 30s baseline segments and matched to the rear-end crashes and near-crashes in the SHRP2SOA8 study \citep{victor2015analysis}, in order to ensure that they were relevant for car-following scenarios. See \citet{victor2015analysis,bargman2024methodological} for more details about baseline extraction and glance annotations.

For an EOFF-based crash causation model, not only is the duration of the glance important, but so is the placement of EOFF glances in relation to the critical event. Following \citet{bargman2024methodological}, we use an EOFF glance anchor of $\tau^{-1}$ of 0.2s$^{-1}$, where $\tau^{-1}$ is the inverse of the optically defined TTC ($\tau^{-1}=\theta/\dot{\theta}$, where $\theta$ is the optical angle of the width of the LV on the driver's retina, and $\dot{\theta}$ its time-derivative). As in \citet{bargman2024methodological}, we assume that a) the LV has an equal probability of braking at any time, b) the FV driver follows the EOFF glance distribution described above until $\tau^{-1}$ reaches 0.2$s^{-1}$, and c) if the FV driver is looking at the road ahead when $\tau^{-1}=0.2s^{-1}$, they will not look away anymore. These assumptions are based on a study of naturalistic braking in critical events reported in \citet{markkula2016farewell}. This results in a glance anchoring scheme where only glances overlapping $\tau^{-1}=0.2s^{-1}$ (just before the crash) are considered relevant for crash causation. Consequently, the original EOFF glance distribution can be transformed into an overshot distribution, which describes the probabilities of off-road glances exceeding the anchor point at $\tau^{-1}=0.2s^{-1}$---hereafter referred to as OEOFF. See \citet{bargman2024methodological} for a description of the overshot distribution. Once the OEOFF is obtained, the crash-causation model assumes that OEOFF glances are simply “placed” with their starting point at $\tau^{-1} = 0.2s^{-1}$.  Note that OEOFF durations range from 0s to 6.6s. For the simulations, we used a bin size of 0.1s, resulting in a total of 67 bins of EOFF and OEOFF glance durations.

\subsubsection{A driver maximum deceleration crash-causation sub-model}
\label{sec:deceleration_crash_causation_model}

The second element of the crash-causation model is the maximum deceleration of the FV driver when there is a critical event. In safety-critical situations, deceleration generally increases quickly until it reaches a maximum. Surprisingly, the literature has shown no clear correlation between the urgency of a situation and maximum deceleration (although urgency is correlated with reaction time and jerk; \citeauthor{markkula2016farewell},2016). That is, drivers do not seem to use the vehicle's or roadway’s full potential, even if they are about to crash. This lack of optimal deceleration by drivers is the second part of our crash-causation model: drivers who could have braked more aggressively, but did not, contribute to crash causation and severity. Consequently, different maximum braking levels are used in the simulations, with a jerk value of approximately 20 m/s$^3$. Specifically, a maximum deceleration distribution was extracted by fitting a piecewise linear model to the pre-crash kinematics of rear-end crashes in the SHPR2 naturalistic driving dataset. For the simulations, discrete decelerations ranging from -10.25 m/s$^2$ to -3.75 m/s$^2$ with bin widths of 0.5 m/s$^2$ were chosen from the fitted deceleration distribution.

\subsubsection{Combining two crash-causation sub-models}
\label{sec:comined_crash_causation_model}

\begin{figure}[htb!]
\begin{center}
\includegraphics[width=1\textwidth]{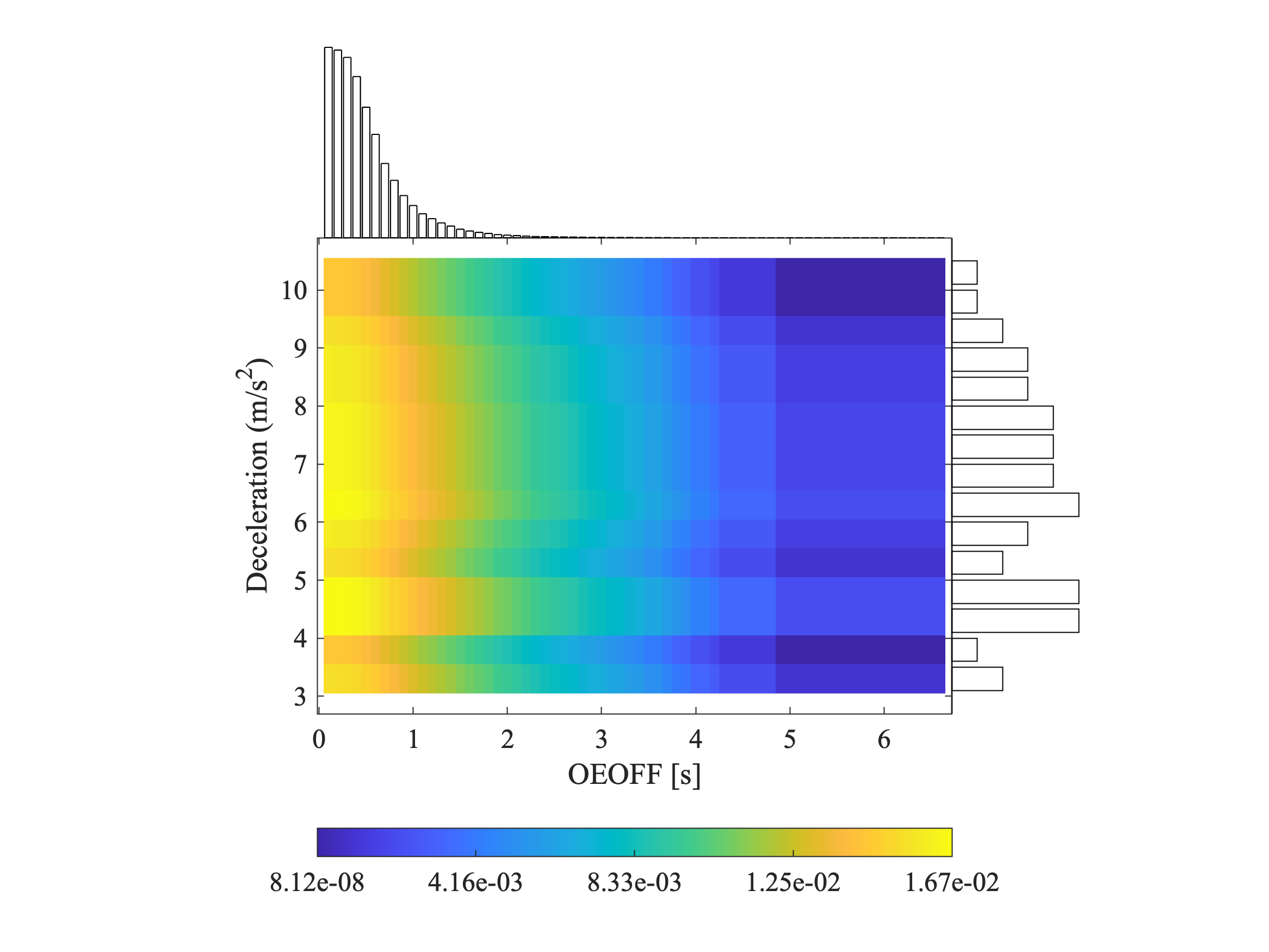}
\end{center}
\caption{Heatmap of the joint probability of OEOFF and deceleration, with their marginal distributions above and to the right, respectively. Glances of 0s are excluded in this plot.}
\label{fig:heatmap_without0s}
\end{figure}

As mentioned, we assume that OEOFF glances and maximum decelerations are independent. That is, the probability of any specific combination of OEOFF duration and maximum driver deceleration is simply the product of the two marginal probability density functions. This joint probability distribution is the core of the crash-causation model, and the joint and marginal distributions are shown in Figure \ref{fig:heatmap_without0s}. Since most glances occur at 0s with probability 0.854, this category is excluded from the glance distribution illustration in Figure \ref{fig:heatmap_without0s}. 

\FloatBarrier

\subsubsection{An advanced driver assistance system}
\label{sec:ADAS}

This paper applied sampling methods to assess a specific safety system and estimate its performance, using counterfactual simulations. The system evaluated is an automated emergency braking (AEB) system provided by Zenseact \citep{zenseact}. The AEB system assessed is a company-internal system used for testing purposes, not a production system. The counterfactual simulations compare the crashes outcomes in baseline simulations without the system (generated using the crash-causation and driver response models) to those in simulations where the AEB is applied, evaluating its safety impact.

\subsubsection{Virtual simulation framework}
\label{sec:virtual_simulation_framework}

The virtual simulations were carried out in the Volvo Car Corporation's virtual simulation tool chain, which integrates a combination of vehicle models (including brakes, chassis, powertrain, and tires, etc.) as well as driver models. The simulations in this study were executed from a Python script. The parameter settings for each prototype event were created by the script, which combined OEOFF durations with maximum deceleration values while keeping the information about the probability of each combination. The script then instantiated simulations in Esmini \citep{esmini}, which loaded and executed the individual concrete scenarios. Specifically, for each simulation, the parameters of the crash-causation models (glance and deceleration inputs) were set. The simulation output was the relative impact speed at impact between the LV and the FV, where a relative impact speed of zero indicated a successfully avoided collision. Injury risk was calculated based on an injury risk function as a function of $\Delta v$ from \citet{stigson2012injury}, for both baseline and countermeasure simulations (i.e., simulations where the safety system was virtually applied to  baseline scenarios). An injury risk function for Maximum Abbreviated Injury Scale equal or greater than two (MAIS2+) was used. The calculation of $\Delta v$ was simplified by assuming that the two vehicles had the same mass, and that the coefficient of restitution was zero (i.e., a perfectly inelastic collision). Therefore, $\Delta v$ was taken as  half of the relative speed at impact (hereafter referred to as impact speed). We then calculated reductions in impact speed and injury risk for each original prototype event and OEOFF-maximum deceleration pair. Finally, the simulation results was exported to $\texttt{R}$ \citep{R_core_team} for empirical evaluation of the adaptive sampling methods.

\subsubsection{The ground truth dataset}
\label{sec:ground_truth}
\FloatBarrier

In this work, a ground-truth dataset was created, and its safety metrics were used to assess the sampling methods. The dataset comprised all generated scenarios based on all possible combinations of parameter values across all prototype events. It was constructed by running virtual simulations for all 1,005 possible pairs of OEOFF durations (67 levels, ranging from 0.0 to 6.6s) and deceleration values (15 levels, from 3.75 to 10.25 m/s$^2$) for each of the 44 prototype events. Each simulation was conducted both under normal driving conditions (baseline scenario) and with the virtual AEB system applied, resulting in a total of 88,440 simulations (Table \ref{tab:ground_truth_table}), same as in \citet{Imberg2024}.

In contrast to \citet{Imberg2024}, this study evaluates the safety impact as the average across all prototype events while ensuring an equal contribution of each prototype event to the overall safety impact assessment. Equal contribution is achieved either through stratification, where the target characteristic is first estimated separately for each prototype event and then averaged across cases, or through post-stratification, where each instance is weighted after each sampling iteration using case-specific weights computed as the inverse of the total OEOFF–deceleration probability within the crash region of that particular case. When applied to the full dataset (ground truth), both approaches are mathematically equivalent and yield identical results.

\begin{figure}[htb!]
\begin{center}
\includegraphics[width=1\textwidth]{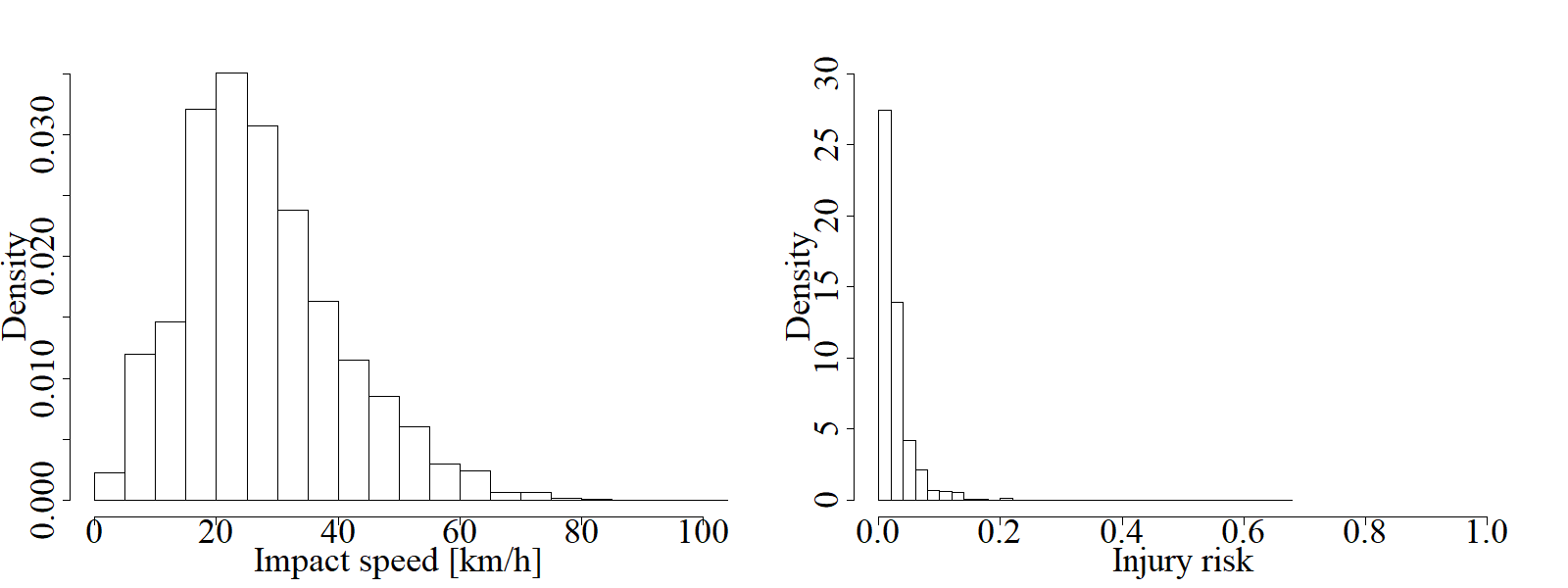}
\end{center}
\caption{Empirical distributions of baseline impact speed (left) and MAIS2+ injury risk (right) in the full ground-truth dataset of simulated crashes.}
\label{fig:baseline_ground_truth}
\end{figure}

The full-grid impact speed and injury risk distributions for the baseline scenarios across all 44 cases are shown in Figure \ref{fig:baseline_ground_truth}. The median (range) of the maximum impact speed across the 44 prototype events was 53.6 (14.9–109.9) km/h. Approximately 38.5\% of baseline simulations resulted in crashes. The ground truth impact speeds and injury risks for the baseline scenarios (excluding non-crashes) are shown in Figure \ref{fig:baseline_ground_truth}, while Table \ref{tab:baseline_and_AEB} presents a comparison between baseline and the countermeasure simulations.

\begin{table}[htb!]
\caption{Parameter levels and values used in the crash causation-based scenario generation in this study. \label{tab:ground_truth_table}}
\small
\centering
\begin{tabular}{p{10.9cm}%
                >{\raggedleft\arraybackslash}p{4.5cm}} 
Parameter & Specification \\ 
\hline
OEOFF glance duration, number of levels (range) & 67 (0.0–6.6 s) \\
Maximum deceleration, number of levels (range) & 15 (3.75–10.25 m/s$^2$) \\
Number of prototype events & 44 \\
Total parameter combinations per prototype event & 1,005 \\
Total number of simulation scenarios (baseline + countermeasure) & 88,440 \\Maximum impact speed per prototype event, median (range) & 53.6 (14.9–109.9 km/h)\\
Proportion of crashes in baseline scenario per prototype event, median (range) & 0.34 (0.0004–1.00)\\
Overall proportion of crashes in baseline scenario & 0.385\\
\hline
\end{tabular}
\end{table}
\FloatBarrier

\FloatBarrier

\begin{table}[htb!]
\caption{Descriptive statistics of simulation outcomes in the crash-causation-based scenario generation, comparing manual driving (baseline scenarios) and countermeasure scenarios with a automated emergency braking (AEB). In countermeasure scenarios, instances where crashes were completely avoided are included in both the impact speed and the injury risk calculations. \label{tab:baseline_and_AEB}}
\centering
\begin{tabular}{lrrr} 
& \multicolumn{1}{c}{Baseline} & AEB & Difference \\ \hline
Mean impact speed (km/h)  & \multicolumn{1}{c}{26.3}  & 3.0   & -23.3  \\ 
Mean MAIS2+ injury risk & \multicolumn{1}{c}{0.027} & 0.002 & -0.025 \\ 
Proportion of crashes        & \multicolumn{1}{c}{100\%}  & 15.5\%& -84.5\%\\ \hline
\end{tabular}
\end{table}
\FloatBarrier

\subsection{Sampling methods}
\label{sec:sampling_methods}

We implemented two versions of importance sampling along with the active sampling proposed by \citet{Imberg2024}. To improve efficiency, we applied ASSR logic with the goal of bypassing scenario sampling and simulation runs when the outcomes could be inferred from previous simulations. Additionally, we evaluated the impact of stratification on sampling efficiency. General implementation details are provided in Section \ref{sec:general_implementation_details}. The importance sampling methods are described in Section \ref{sec:importance_sampling}, followed by descriptions of ASSR logic, stratification, post-stratification, and active sampling in Sections \ref{sec:use_logic} to \ref{sec:active_sampling}.

\subsubsection{General implementation}
\label{sec:general_implementation_details}

All sampling methods were implemented iteratively, selecting simulation scenarios (i.e., OEOFF and maximum deceleration parameter values) in batches according to a multinomial distribution. Here, a batch refers to the number of samples chosen in one iteration. For importance sampling, the sampling probabilities within the joint distribution of OEOFF and maximum deceleration remained fixed throughout the experiment, whereas for active sampling, they were updated iteratively. 

\subsubsection{Importance sampling methods}
\label{sec:importance_sampling}

We implemented two importance sampling schemes: density importance sampling and severity importance sampling. Density importance sampling assigns sampling probabilities proportional to the probability of the OEOFF–deceleration pair. This approach aligns with the scenario generation framework, which prioritizes instances with high OEOFF–deceleration probabilities, as these contribute more to the estimation of key outcome variables, including the number of crashes, impact speed, and injury risk.

Severity importance sampling aims to further reduce variance in safety impact evaluation by oversampling high-severity instances. Given that the safety impact of an AEB system increases with impact severity, this scheme assigns sampling probabilities proportional to $w_i \times o_i \times  d_i \times m_i$, normalized to sum to $1$. Here $w_i$ represents the probability of the OEOFF–deceleration pair for instance $i$, $o_i$ is the corresponding off-road glance duration, $d_i$ is the maximum deceleration, and $m_i$ is an \textit{a priori} known maximum possible impact speed of instance $i$. To account for differences in variable scaling, all continuous variables (off-road glance duration, deceleration, and maximum impact speed) were normalized to the range $[0.1, 1]$ before computing the severity sampling probabilities. 

To determine the maximum impact speed for each prototype event, severity sampling was initialized with a deterministic sample of 44 baseline-scenario simulations (one per prototype event), observed at a maximum glance duration of 6.6s and a minimal deceleration of 3.3 $m/s^2$ (see Section \ref{sec:use_logic} for further details).

\subsubsection{Using ASSR logic to reduce sampling space}
\label{sec:use_logic}

In this application, crash severity is monotonically related to both glance duration and deceleration level. Some scenarios may not even result in a crash---for instance, when glance duration is short and/or braking is sufficiently strong. For a given deceleration level, as OEOFF duration increases from 0s to 6.6s, there exists a threshold at which the first crash occurs. Similarly, as glance duration further increases, there may come a point where it is too late for the driver to react at all, meaning that under the given response model, the crash happens before any braking is applied. Beyond this point, for all longer glance durations, the impact speed of crashes remains constant, as no braking occurs. Consequently, large regions of the simulation parameter space may consist entirely of non-crashes or crashes at maximum impact speed. Based on this structure, we can deduce the following:

\begin{enumerate}[label=\roman*)]
    \item \textbf{Eliminating unnecessary baseline simulations}: If a non-crash is observed in the baseline scenario for a specific prototype event, off-road glance duration, and deceleration level, then all less severe variations of that case (i.e., shorter OEOFF durations and larger decelerations) will also not result in a crash. Since our interest lies in assessing the system's effectiveness in avoiding or mitigating baseline crashes, simulations in such regions do not need to be conducted. 
    \item \textbf{Skipping countermeasure simulations for non-crash scenarios}: A simulation with a countermeasure (e.g., an AEB system) will never result in a crash if the corresponding baseline scenario (without the countermeasure) did not result in a crash. Therefore, baseline simulations should be executed first, and countermeasure simulations only need to be run for instances where a crash occurs in the baseline scenario.
    \item \textbf{Reducing countermeasure simulations for mitigated crashes}: If a crash is observed in the baseline scenario but does not occur with the countermeasure for a given prototype event, OEOFF duration, and deceleration level, then we know that none of the less severe variations (i.e., shorter OEOFF duration and greater deceleration) with the countermeasure will produce a crash. Thus, it suffices to run simulations only for the baseline scenario in that sampling space.
    \item \textbf{Inferring impact speed in extreme scenarios}: Each prototype event has a maximum impact speed, which occurs in the baseline scenario when the OEOFF duration is at its maximum and the deceleration is at its minimum. If a collision at maximum impact speed is observed at a shorter OEOFF duration or greater deceleration, then all more extreme variations (i.e., longer OEOFF durations or smaller decelerations) will also result in collisions at the same impact speed. Consequently, the impact speed in these regions can be inferred without running additional simulations.  
    \label{list:maximal_impact_speed}
\end{enumerate}

By sampling simulation scenarios in small iterative batches, we can progressively exclude regions from the sampling space where non-crashes are observed in the baseline scenario. This approach minimizes unnecessary computations by leveraging previously run simulations to deduce outcomes, thereby improving efficiency in the sampling process.

\subsubsection{Stratification and post-stratification}
\label{sec:stratification}
In this study, both stratification and post-stratification were applied to ensure balanced contributions from prototype events in the overall safety impact assessment and to correct biases arising from variations in crash frequencies across scenarios.

Stratification ensures balance at the sampling stage by dividing the parameter space into predefined strata and allocating samples proportionally. Without stratification, prototype events with a higher crash likelihood in high OEOFF–deceleration probability regions would disproportionately influence the results. By ensuring equal representation, stratification ensures that each prototype event contributes equally to the assessment, regardless of its inherent crash frequency. Estimation was performed by computing the target characteristic separately for each prototype event and then averaging across cases.

Post-stratification, in contrast, applies post hoc reweighting \citep{franco2017developing} when sampling is not explicitly stratified. Since crash frequencies and associated probabilities vary across the prototype events, each instance is assigned a weight equal to the inverse of the total OEOFF–deceleration probability within its crash region, ensuring proportional representation in the final estimates.

\subsubsection{Active sampling}
\label{sec:active_sampling}

\begin{figure}[!htb]
\centering
\captionsetup{skip=10pt} 
\includegraphics[width=0.8\textwidth]{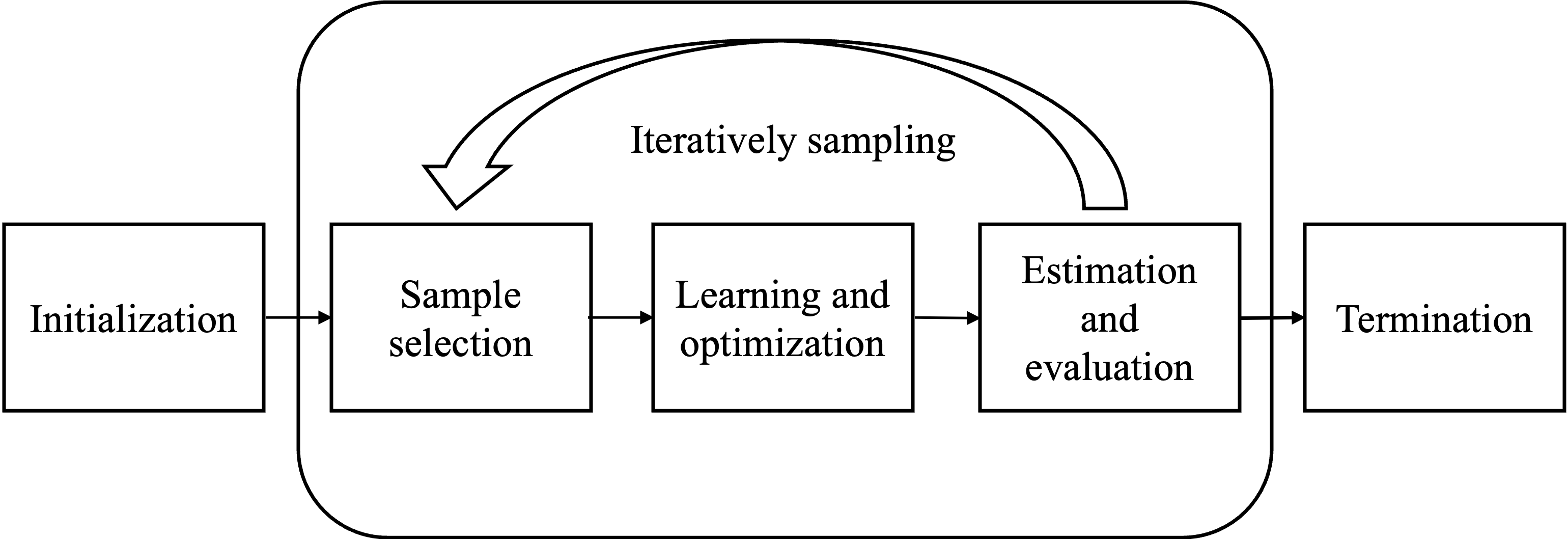}
\caption{Flowchart illustrating the active sampling process.}
\label{fig:active_sampling}
\end{figure}

The active sampling method, introduced by \citet{Imberg2024}, is illustrated in Figure \ref{fig:active_sampling}. Unlike traditional 'passive' importance sampling methods, active sampling employs machine learning to iteratively optimize the sampling scheme, offering an adaptive and data-driven approach to sample selection. Below, we provide a brief overview of the active sampling method and its implementation in this study.

\begin{enumerate}[label = \roman*)]
    \item \textbf{Input}: The input to the algorithm consisted of a dataset of potential simulation scenarios, defined by prototype events and combinations of OEOFF duration and deceleration. The algorithm was optimized for a target characteristic, such as the mean impact speed reduction, crash avoidance rate, or mean injury risk reduction. Additional input parameters included batch size $n_t$, maximum number of iterations $T$, and a target precision $\delta$, which determined the stopping condition.
    \item \textbf{Initialization}: The active sampling algorithm was initialized with a deterministic sample of 44 instances, corresponding to prototype events with the maximum OEOFF duration (6.6s) and minimum deceleration (3.3m/s$^2$). This allowed the maximum impact speed for each prototype event to be deduced (see Section \ref{sec:use_logic}), since it was expected to be an important predictor of the case-specific safety impact profile. Consequently, maximum impact speed was incorporated into subsequent learning and optimization steps to enhance predictive accuracy during the learning phase and improve the overall efficiency of the sampling algorithm.
    \item \textbf{Learning}: Observed data were used to train models predicting the probability of collision in the baseline scenarios, as well as the expected impact speed reduction, injury risk reduction, and probability of collision in the countermeasure scenarios, depending on the target characteristic. Modeling was performed using the random forest method \citep{Breiman2001}, implemented via the \texttt{ranger} package (version 0.14.1) \citep{ranger}, with hyperparameter tuning conducted through cross-validation using the \texttt{caret} package (version 6.0-92) \citep{caret} in R \citep{R_core_team}. The explanatory variables included OEOFF duration, maximum deceleration, and case-specific maximum impact speed. The learning step was implemented from the second iteration onward. 
    \item \textbf{Optimization}: Based on the trained models, optimal sampling probabilities were computed according to Equation \eqref{eq:optimal_sampling_scheme} in Section \ref{sec:optimal_sampling_probabilities}. If the learning step failed due to insufficient data (e.g., no crashes were generated) or if the machine learning model's accuracy fell below a pre-defined threshold (e.g., R-squared $<$0 or classification accuracy $<$0 on hold-out data---a portion of the dataset reserved for validation and not used during training; \citeauthor{james2013introduction}), density importance sampling was used as a fallback instead of active sampling. This fallback approach is theoretically optimal in the active sampling algorithm when no auxiliary information is available. 
    \item \textbf{Sampling}: A batch of $n_t$ new simulation scenarios was selected at random according to a multinomial sampling design. Both baseline scenarios and corresponding countermeasure scenarios were executed. Simulation outputs included impact speed in both baseline and countermeasure scenarios, as well as  the calculated outcomes: impact speed reduction, injury risk reduction, and crash avoidance with the countermeasure.
    \item \textbf{Estimation}: The target characteristics were estimated using inverse probability weighting. Additionally, the standard error of the current estimate was calculated to assess its precision. Methods for parameter and variance estimation were detailed in \citet{Imberg2024}, with adjustments for stratification and post-stratification outlined in Section \ref{sec:stratification} above.
    \item \textbf{Termination}: The algorithm terminated if the standard error of the current estimate of the target characteristic (i.e., mean impact speed reduction, mean injury risk reduction, or crash avoidance rate) fell below the pre-determined target precision $\delta$ or the maximum number of iterations $T$ was reached. If neither condition was met, the process returned to the Learning step for another iteration.  
\end{enumerate}

\subsubsection{Optimal sampling probabilities}
\label{sec:optimal_sampling_probabilities}

The optimal sampling probabilities are given by 
\begin{gather}
\label{eq:optimal_sampling_scheme}
   \pi_i \propto c_i, \\
   \nonumber
    c_i = \sqrt{\hat p_iw_i^2 \left[(\hat y_i - \hat {\mu})^2 + \sigma^2\right]}. 
\end{gather}
Here, $i$ represents the index of a simulation scenario (baseline-countermeasure pair), which is defined by a combination of prototype event, OEOFF duration, and deceleration. The quantity $\hat{p}_i$ denotes the predicted probability of a collision in the baseline scenario. The term $w_i$ represents the probability of the simulation scenario $i$, as determined by the crash-causation model described in Section \ref{sec:deceleration_crash_causation_model}, while $\hat y_i$ represents the predicted outcome. The predicted outcome may correspond to impact speed reduction, injury risk reduction (with the countermeasure compared to manual baseline driving), or the probability of collision in the countermeasure scenario. The term $\hat{\mu}$ is the estimated mean outcome, which may represent the mean impact speed reduction, mean injury risk reduction, or crash avoidance rate, depending on the target characteristic of interest. Finally, $\sigma$ represents the residual standard deviation or root mean squared error (RMSE) of the corresponding prediction model. For a continuous outcome, $\sigma$ was taken as the RMSE on hold-out data, whereas for a binary outcome $y_i$, it was defined as $\sigma = \sqrt{\hat y_i (1- \hat y_i)}$. A formal proof and derivation of this formulation can be found in \citet{Imberg2024}.

In the stratified setting, the target characteristics and sampling probabilities given by Equation \eqref{eq:optimal_sampling_scheme} were evaluated separately for each prototype event, with the sampling probabilities normalized within each case to sum to 1. The prediction models, however, were trained across all cases to maximize the use of available data. To stabilize performance in case-specific estimation of the mean impact speed reduction, mean injury risk reduction, or crash avoidance rate in small samples, a Bayesian-inspired shrinkage procedure was employed. Each case-specific estimate $\hat \mu_k$ was adjusted using a weighted combination of the observed case-level characteristic and the overall characteristic, with increasing weight assigned to the observed data and decreasing weight assigned to the grand mean as the number of generated crashes per case increased:
\[
\hat \mu_k^{\text{shrunk}} = \rho_k \hat \mu_k + (1 - \rho_k)\bar \mu, \quad \rho_k = \frac{\sigma^2_u}{\sigma^2_u + \sigma^2_k/n_k}
\]
where $\hat \mu_k$ is the case-specific safety impact characteristic, and $\bar \mu$ represents the grand mean or the combined target characteristic across all cases. The quantity $\sigma^2_u$ represents the variance of the estimated target characteristic across cases, $\sigma^2_k$ the variance within case $k$, and $n_k$ is the sample size (i.e., the number of crashes in the baseline scenario) for case $k$. If no crashes were generated for a specific case, $\rho_k$ was set to $0$, ensuring that the estimate defaults to the overall mean. This approach stabilizes estimates for cases with limited data while allowing those with sufficient observations to remain closer to their empirical values, mimicking the maximum a posteriori (MAP) estimate of the mean in a Bayesian normal model.

In the non-stratified setting, sampling probabilities were calculated similarly to Equation \eqref{eq:optimal_sampling_scheme} but with the values of $c_i$ for scenarios $i$ corresponding to a common prototype event $k$ pre-multiplied by a case-specific normalization factor $u_k = 1 / \sum_{j} \hat{p}_j w_j$ before calculating the sampling probabilities, where the summation over $j$ includes all indices corresponding to case $k$. As above, $\hat{p}_j$ denotes the predicted probability of a collision in the baseline scenario. This adjustment accounts for the re-weighting procedure used in estimation, ensuring that each prototype event contributes equally to the overall estimation. Sampling was performed over the entire input space, with probabilities normalized to sum to 1 across all prototype events.

\subsection{Stopping conditions}
   \label{sec:stopping conditions for sampling}
To conserve resources, a stopping condition is required to terminate sampling once sufficient data have been collected. By monitoring the confidence interval width---or equivalently, the standard error of the estimate---we can determine when the desired precision has been reached. When the confidence interval becomes sufficiently narrow, or the standard error falls below a predefined threshold, sampling can be terminated.

Stopping conditions are typically informed by domain knowledge and may be based on metrics such as the region of practical equivalence (ROPE) or the coefficient of variation \citep{schwaferts2020bayesian, kukukova2008impact,carrasco2003estimating}. The appropriate thresholds for these metrics should be chosen based on expert judgment. A ROPE value defines the range within which differences are considered practically equivalent and can be determined through, for example, expert consensus in a workshop setting. For instance, if experts agree that a crash avoidance rate within $\pm$ 0.05 (five percentage points) is practically equivalent, then a threshold of 0.025 for the standard error or 0.05 for the confidence interval half-width can be chosen as the stopping threshold. In contrast, the coefficient of variation is a percentage representing the allowable uncertainty of the estimate relative to the quantity of interest. A typical value, such as 2.5\%, ensures that the standard error does not exceed 2.5\% of the estimated parameter value, providing a relative measure of precision. Thus, the stopping condition is met when the standard error drops below this percentage of the target characteristic estimate.

Another practical stopping criterion is the limitation imposed by available resources, such as computational capacity or constraints. If only a fixed proportion, for example 10\% of the total possible simulations, can be run due to resource limitations, sampling must stop once this limit is reached. Similarly, if there is a fixed amount of CPU time, memory or wall-clock time allocated, sampling must conclude when the allocated resources are exhausted.

\subsection{Metrics and simulation performance evaluation}
\label{sec:performance_evaluation}

Sampling performance was evaluated using the RMSE of the estimator for the mean impact speed reduction, crash avoidance rate, or mean injury risk reduction. RMSE was calculated as the square root of the mean squared deviation of the estimate from the ground truth across 200 independent subsampling experiments. This was analyzed and presented graphically as a function of the number of virtual crash-causation model simulations. Unless otherwise stated, the batch size was set to ten simulations per iteration. 

The following comparisons were performed:
\begin{enumerate}[label = \roman*)]
    \item \textbf{Comparison of sampling methods}: Active sampling compared to density importance sampling and severity importance sampling.
    \label{list:average targets}
    \item \textbf{Effect of domain knowledge}: Comparison of active sampling with and without domain-knowledge-based ASSR logic.
    \label{list:logic}
    \item \textbf{Stratification versus post-stratification (without ASSR)}: Active sampling and the better-performing importance sampling method (i.e., severity sampling), both without ASSR, comparing stratification to post-stratification.    
\label{list:stratification, without ASSR}
    \item \textbf{Stratification versus post-stratification (with ASSR)}: Active sampling and the better-performing importance sampling method (i.e., severity sampling), both with ASSR, comparing stratification to post-stratification.   
    \label{list:stratification, with ASSR}
    \item \textbf{Impact of batch size}: Active sampling with batch sizes of $44$, $132$, and $440$ per iteration (i.e., 1x, 3x, and 10x the 44 prototype events).
    \label{list:batch_size}

\end{enumerate}

\section{Results}
\label{sec:results}

This section presents the results of the sampling strategy evaluations, starting with a comparison of active sampling and importance sampling in Section \ref{sec:results_caseweighting}, followed by the effect of ASSR logic in Section \ref{sec:results:active_sampling_with_withoutlogic}, stratification versus post-stratification in \ref{sec:results_stratification}, and the influence of batch size on active sampling performance in Section \ref{sec:results_batching}.

\subsection{Comparison of importance sampling and active sampling performance}
\label{sec:results_caseweighting}

\begin{figure}[htb!]
\begin{center}
\includegraphics[width=0.9\textwidth]{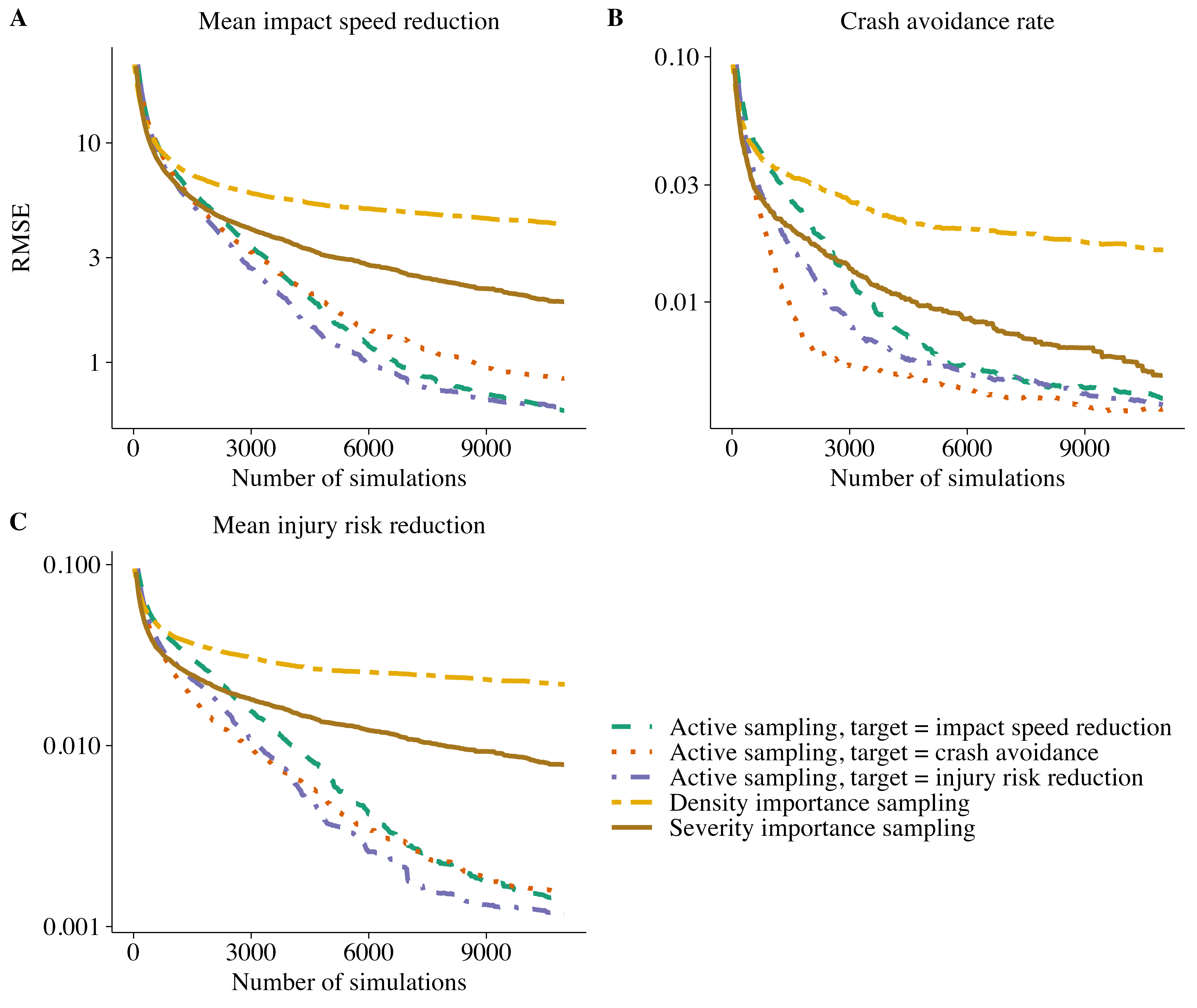}
\end{center}
\caption{Root mean squared error (RMSE) for estimating (A) mean impact speed reduction, (B) crash avoidance rate, and (C) mean injury risk reduction using active sampling, compared to density and severity importance sampling. Active sampling consistently outperformed importance sampling, except for small sample sizes, where their performance was comparable. Post-stratification (case weighting) was applied to ensure equal contribution from all prototype events.}
\label{fig:nonstratify_withoutlogic}
\end{figure}

The RMSE of active sampling is compared with density and severity importance sampling (Figure \ref{fig:nonstratify_withoutlogic}). For this comparison, post-stratification (case weighting) was used, ensuring that each prototype event contributed equally to the estimation. Active sampling outperformed importance sampling for all three targets---mean impact speed reduction, crash avoidance rate, and injury risk reduction\textemdash regardless of the optimization target. However, for small samples, the methods performed similarly. As expected, active sampling performed best for the specific target characteristic it was optimized on.

\FloatBarrier

\subsection{Effect of ASSR logic on active sampling performance} 
\label{sec:results:active_sampling_with_withoutlogic}

\begin{figure}[htb!]
\begin{center}
\includegraphics[width=0.9\textwidth]{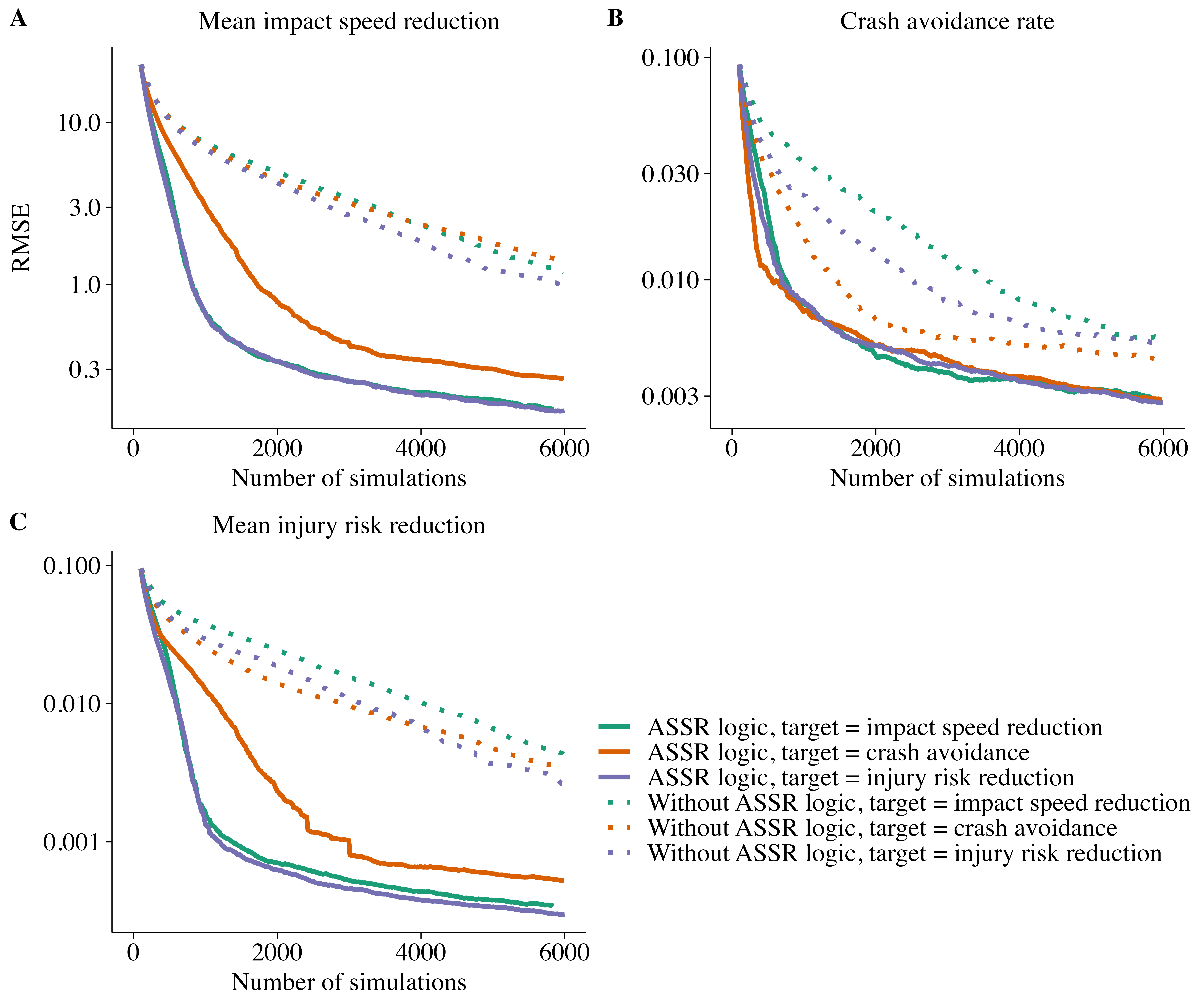}
\end{center}
\caption{Root mean squared error (RMSE) for estimating (A) mean impact speed reduction, (B) crash avoidance rate, and (C) mean injury risk reduction using active sampling with and without ASSR logic. ASSR logic provided substantial performance improvements across all settings.}
\label{fig:actitve_with_without_logic}
\end{figure}

The RMSE of active sampling with and without domain-knowledge-based ASSR logic are compared in Figure \ref{fig:actitve_with_without_logic}. For all three target characteristics, sampling with ASSR logic outperformed sampling without it for the same number of simulations. At 6,000 simulations, for example, the RMSE was reduced by 32\% to 89\%, depending on the outcome.

\FloatBarrier

\subsection{Effect of stratification vs post-stratification on sampling performance}
\label{sec:results_stratification}

\begin{figure}[htb!]
\begin{center}
\includegraphics[width=0.9\textwidth]{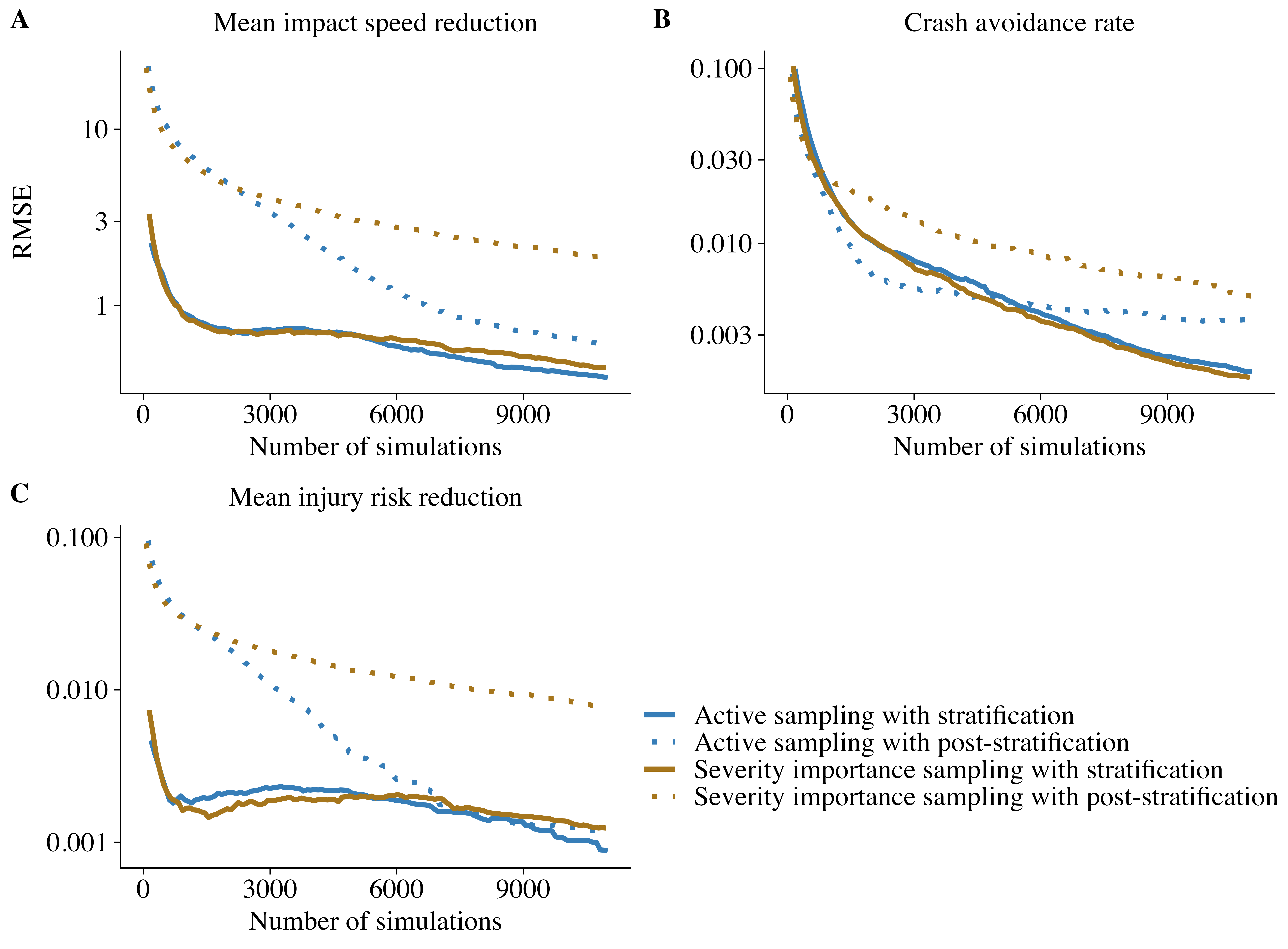}
\end{center}
\caption{Root mean squared error (RMSE) for estimating (A) mean impact speed reduction, (B) crash avoidance rate, and (C) mean injury risk reduction using active sampling and severity importance sampling, each with either stratification or post-stratification. Neither method used ASSR logic. Active sampling was optimized for mean impact speed reduction in (A), crash avoidance rate in (B), and mean injury risk reduction in (C). Stratification improved performance for both methods, while active sampling with post-stratification approached the performance of stratification as sample size increased.}
\label{fig:stratification_withoutlogic}
\end{figure}

Figure \ref{fig:stratification_withoutlogic} compares the RMSE of active sampling and severity importance sampling with and without stratification, excluding ASSR logic. Both active sampling and severity importance sampling generally performed better with stratification than with post-stratification. Without ASSR logic or stratification, active sampling outperformed severity importance sampling. However, when stratification was applied, their performances were comparable. Moreover, as the sample size increased, the performance of active sampling with post-stratification approached that of stratification.

\begin{figure}[htb!]
\begin{center}
\includegraphics[width=0.9\textwidth]{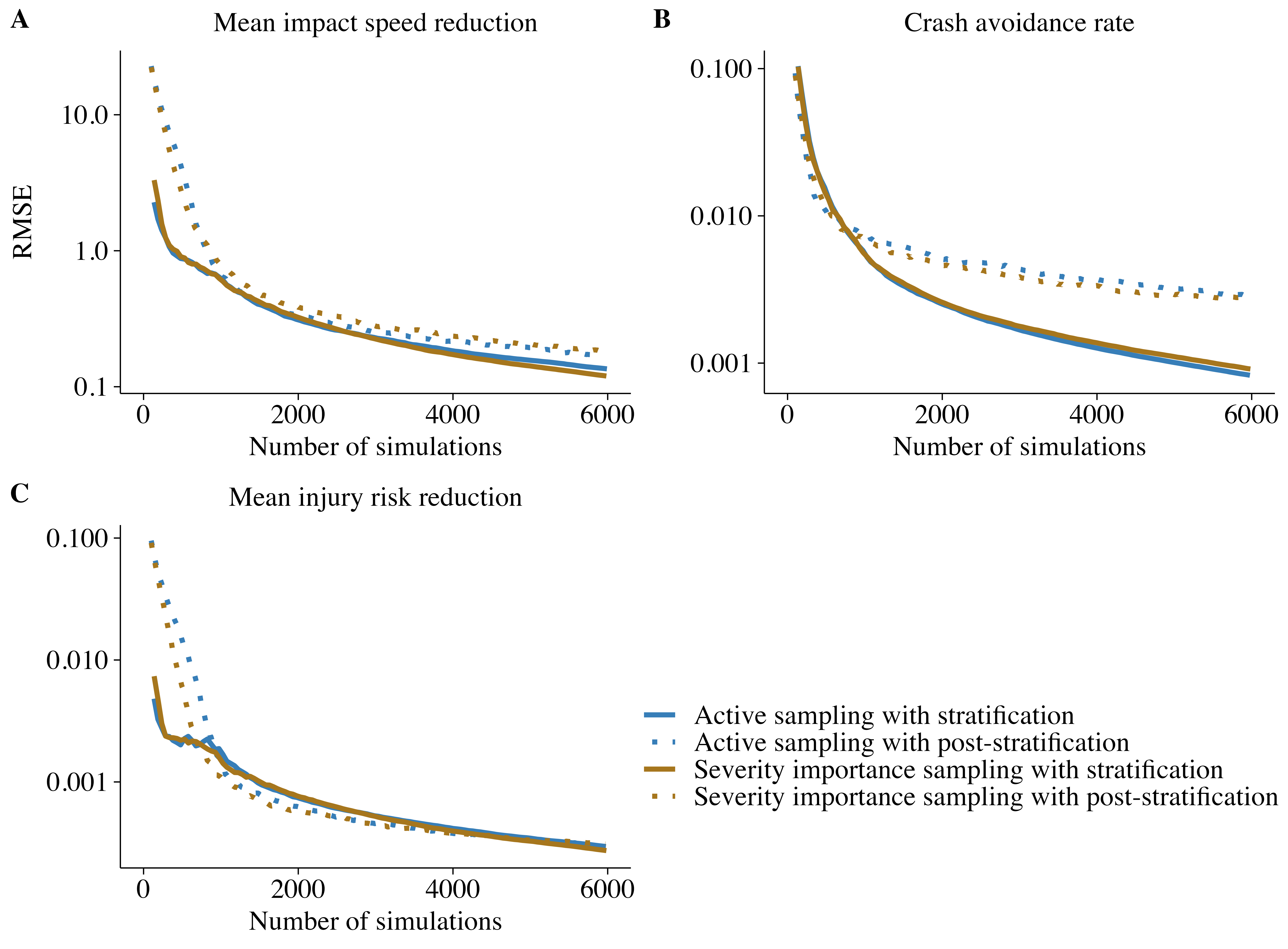}
\end{center}
\caption{Root mean squared error (RMSE) for estimating (A) mean impact speed reduction, (B) crash avoidance rate, and (C) injury risk reduction using active sampling and severity importance sampling, each with either stratification or post-stratification. Both methods incorporated ASSR logic. Active sampling was optimized for mean impact speed reduction in (A), crash avoidance rate in (B), and mean injury risk reduction in (C). With ASSR logic was included, active sampling and severity importance sampling had similar performance using both stratification and post-stratification, although stratification improved performance for crash avoidance rate.}
\label{fig:stratification_withlogic}
\end{figure}

Figure \ref{fig:stratification_withlogic} compares the RMSE of active sampling and severity importance sampling (the better-performing importance sampling method) using stratification versus post-stratification, this time incorporating ASSR logic. For mean impact speed reduction and injury risk reduction, stratification and post-stratification yielded similar results. However, stratification improved performance for crash avoidance rate. Additionally, when either ASSR logic or stratification were incorporated, active sampling and severity importance sampling performed similarly (shown in Figure \ref{fig:stratification_withoutlogic}, \ref{fig:stratification_withlogic}).

\FloatBarrier

\FloatBarrier

\subsection{Effect of batch size on active sampling performance}
\label{sec:results_batching}
\begin{figure}[htb!]
\begin{center}
\includegraphics[width=0.9\textwidth]{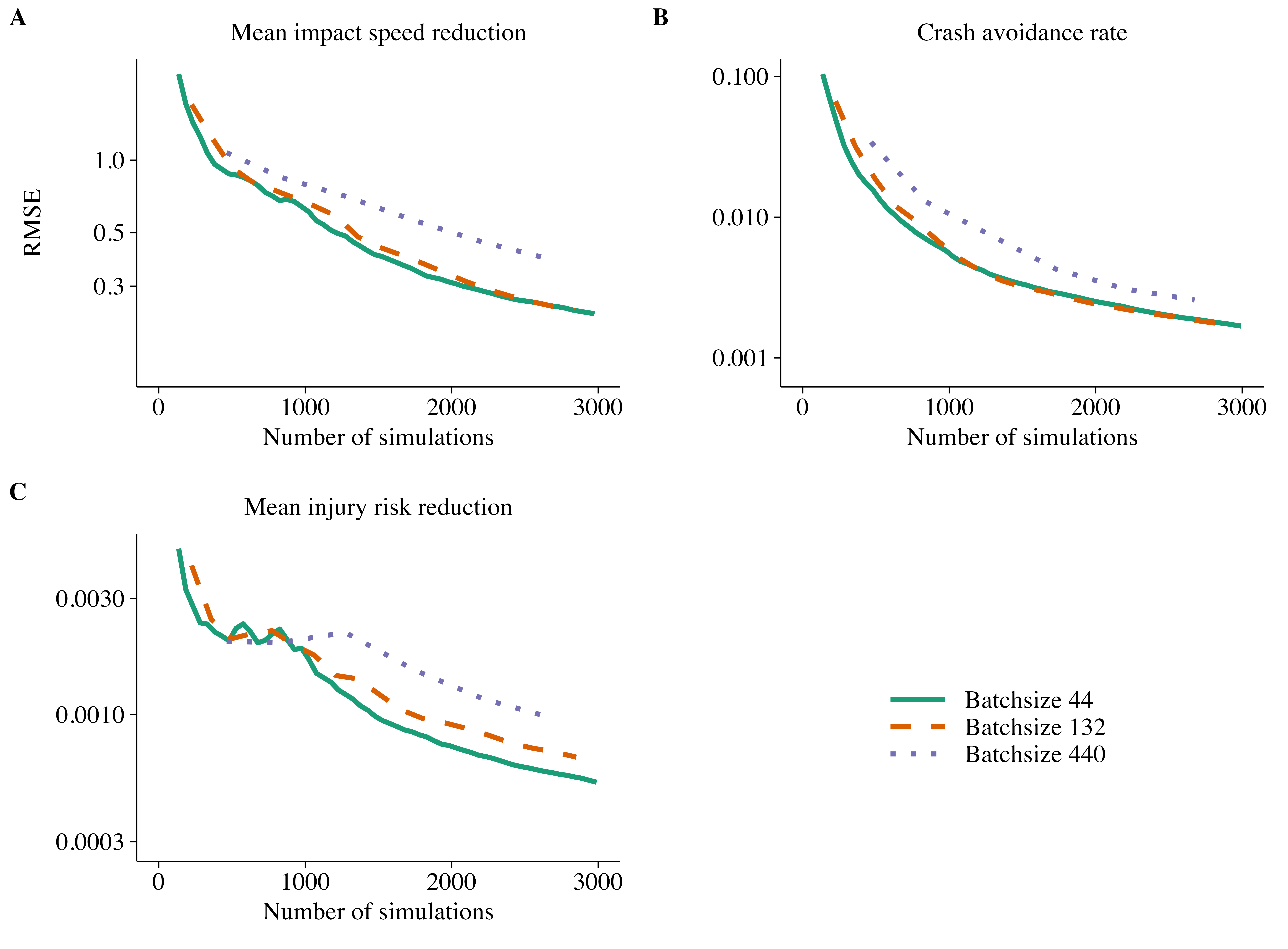}
\end{center}
\caption{Root mean squared error (RMSE) for estimating (A) mean impact speed reduction, (B) crash avoidance rate, and (C) mean injury risk reduction using active sampling with batch sizes of 44 (one per prototype event), 132 (three per prototype event), or 440 (ten per prototype event) per iteration in the active sampling algorithm. Both ASSR logic and stratification were used. Active sampling was optimized for mean impact speed reduction in (A), crash avoidance rate in (B), and mean injury risk reduction in (C). Performance gradually declined as batch size increased.}
\label{fig:batch_size}
\end{figure}

Figure \ref{fig:batch_size} shows the RMSE of active sampling (with ASSR logic and stratification) for varying batch sizes. Performance declined as the batch size increased, but the effect was gradual. There were minor performance losses when increasing from 44 (one per prototype event) to 132 (three per prototype event), but more noticeable reductions as batch size further increased to 440 (ten per prototype event).

\FloatBarrier

\section{Discussion}
\label{sec:discussion}

This study advances the methodology for adaptive sampling in scenario-generation-based safety system assessment by refining existing approaches and integrating domain knowledge to enhance efficiency. We examined traditional importance sampling methods and adaptive importance sampling through active sampling, evaluating the impact of three key implementation features: adaptive sample space reduction (ASSR), stratification versus post-stratification, and batch size effects. Our findings indicate that ASSR improved efficiency for both importance sampling and active sampling, with importance sampling benefiting the most. Stratification consistently improved performance, while post-stratification yielded comparable results to stratification for active sampling at larger sample sizes. Active sampling outperformed importance sampling when no domain knowledge was incorporated, but their performance was similar when ASSR and stratification were applied. The following discussion explores these results, their broader implications, and potential directions for future research.

Compared to previous work on active sampling in crash-causation-based scenario generation by \citet{Imberg2024}, this study enforced a balanced contribution from the prototype events to the target characteristic of interest. In the context of safety impact assessment, crash exposure must be explicitly accounted for. Since prototype events represent crash exposure, each event should contribute equally to the overall safety impact assessment. Weighting is widely used to ensure the representativeness of the crash exposure in datasets \citep{babisch2023leveraging,pfeiffer2007statistical,clark2013inverse}. Consequently, this study employed case weighting—either by design through stratification or by estimation through post-stratification---to ensure that each prototype event made an equal contribution to the estimation of target characteristics. When ASSR and stratification were not applied, our results align with \citet{Imberg2024}, confirming the superiority of active sampling over importance sampling for scenario generation when domain knowledge is limited.

One of the key challenges in implementing importance sampling for safety impact assessment is selecting an appropriate proposal distribution for generating scenarios, as performance depends heavily on how closely the proposal distribution aligns with the ground truth distribution \citep{bugallo2017adaptive}. A poorly chosen proposal distribution may reduce performance \citep{swiler2010importance}, potentially performing even worse than simple random sampling. Consequently, the two importance sampling methods considered in this study had different performance, and while severity-based sampling might be expected to perform better, its advantage cannot be determined in advance. In contrast, active sampling does not rely on a predefined proposal distribution but instead learns the optimal sampling scheme dynamically during the simulation process. However, it does require a target characteristic for optimization, which can influence performance. Extensions of active sampling to multivariate settings, where multiple characteristics are optimized simultaneously, are possible using optimal design theory \citep{Pukelsheim1993, imberg2022, imberg2023}. When the target characteristics of interest are highly correlated, as in this study and often in scenario-generation for safety impact assessment, this has a relatively small impact on sampling efficiency. Ultimately, active sampling provides a more data-driven and robust alternative to importance sampling, particularly in settings where defining an optimal proposal distribution is challenging.

In this work, domain knowledge was incorporated in the form of ASSR, which is closely tied to the crash-causation model that defines crash mechanisms through parameters influencing crash occurrence \citep{davis2011outline}. ASSR enabled the elimination of unnecessary simulations by applying logical constraints: i) longer glances and lower decelerations increase crash risk and impact speed, while ii) shorter glances and greater decelerations increase the likelihood of crash avoidance. These constraints streamline the sampling process by avoiding redundant simulations. More broadly, this approach may be applicable to other scenario-generation problems where the relationship between input parameters and crash severity is known, as more extreme inputs typically correspond to more and more severe crashes, while less extreme inputs lead to milder outcomes. Such structured constraints can improve efficiency, particularly in knowledge-based scenario generation \citep{ding2023survey,da2024ontology,mcduff2022causalcity}, which often relies on predefined expert rules or integrates external knowledge.

When ASSR logic was incorporated, the performance of all sampling methods improved, although the magnitude of the improvement varied across settings. ASSR led to greater efficiency gains for importance sampling, ultimately resulting in equal performance between active sampling and importance sampling when ASSR was applied. This may be due to the partially overlapping objectives and benefits of active sampling and ASSR. Active sampling, guided by machine learning, identifies and oversamples regions with a high likelihood of generating informative scenarios (e.g., with a high likelihood of a crash in the baseline condition). However, if a rule-based approach like ASSR can achieve similar outcomes by systematically reducing the sample space, the added value of active sampling diminishes, explaining its comparatively smaller performance gains.

Stratification is a well-established variance reduction technique, particularly effective in settings with substantial heterogeneity across strata \citep{singh1996stratified, park2024strata, jing2015stratified}, which is also confirmed in this study. Across all settings, stratification significantly improved sampling performance by balancing case representation and ensuring more accurate estimation for all prototype events. However, for injury-risk reduction, RMSE remained unchanged over a substantial range of simulation counts. This may be attributed to the large non-crash or low-severity crash regions in the sampling space, combined with the exponential response curve of the MAIS2+ injury risk function, which required the generation of a sufficient number of high-severity crashes for accurate assessment. Notably, active sampling without stratification approached the performance of stratified sampling as sample sizes increased, as the algorithm dynamically learned the underlying structure over iterations. In contrast, importance sampling consistently exhibited a performance gap between stratified and non-stratified sampling, emphasizing its reliance on explicit stratification to achieve balanced case representation and reinforcing the advantage of active sampling.

Batch size plays a crucial role in both sampling performance and computational efficiency \citep{citovsky2021batch}. In our study, larger batch sizes slowed the adaptive learning process of active sampling, as the algorithm had fewer opportunities to adjust between iterations, leading to reduced sampling efficiency. However, in parallel computing environments, larger batch sizes can significantly reduce wall-clock time, enabling faster overall safety impact assessments. This trade-off highlights the need to balance simulation effort with computational efficiency when selecting batch sizes. Smaller batches are preferable in settings where minimizing the number of simulations is a priority, such as in resource-constrained settings. Conversely, larger batches are advantageous when parallelization resources are available, allowing for faster completion times despite the increased number of simulations required. An alternative to fixed batch sizes is the use of adaptive batch sizes, which dynamically adjust over iterations to balance sampling efficiency and computational performance \citep{ma2021adaptive}. The relationship between batch size and machine learning model learning rate can also be leveraged to scale batch size adaptively across iterations \citep{devarakonda2017adabatch,balles2016coupling}. Balancing batch size considerations ensures optimal performance tailored to the specific computational and resource constraints of the application.

\subsection{Generalization and implications}

The findings of this study have broader implications for improving computational efficiency in scenario generation. ASSR techniques informed by domain knowledge proved highly effective in reducing computational costs and can be extended to other parameterized scenario generation tasks. When such domain knowledge is available, importance sampling with ASSR is highly efficient, whereas in its absence, active sampling remains more efficient. Both methods show benefit from stratification, where applicable, further enhancing performance. Future research could explore extensions of this work to more complex scenarios, including higher-dimensional parameter spaces, continuous sampling spaces, and advanced crash-causation models, to further refine and optimize adaptive sampling techniques for safety impact assessments.

\subsection{Limitations}
\label{sec:Limitations}
This study focused on estimating finite population characteristics, such as means or ratios, in the context of AEB system safety impact assessment. While the proposed methods demonstrated high efficiency for these objectives, they may be less effective for other scenario generation tasks, such as identifying corner cases or extreme scenarios. In such cases, methods designed for anomaly detection \citep{chandola2009anomaly} or Bayesian optimization \citep{frazier2018tutorial}, may be more suitable.

Another limitation concerns the use of domain-knowledge-based logic components within the ASSR technique. While effective for parameterized scenarios, integrating logic becomes more challenging when machine learning methods generate concrete scenarios directly from data without parametrization, as in \citet{zorin2024new}. The lower interpretability of machine learning outputs and the lack of explicit parametrization make it difficult to apply rule-based constraints directly linked to crash outcomes. This restricts the applicability of ASSR and logic-based sampling techniques in settings where explicitly parametrized input data are unavailable.

Finally, this study employed a simplified delta-v calculation, assuming equal vehicle masses and no elasticity during impact. While this simplification reduces the realism of absolute injury risk estimates, it does not affect the validity of the study's conclusions regarding the efficiency of the sampling methods.

\subsection{Conclusion}
\label{sec:conclusion}

This study evaluated how three key implementation features---ASSR, stratification, and batch sampling—influence the efficiency of importance sampling and active sampling for scenario generation in virtual safety impact assessment. Both ASSR and stratification substantially improved efficiency for both methods, with active sampling performing better in the absence of these features, while importance sampling performed on par with active sampling when either stratification or ASSR was applied. Batch size influenced performance, with moderate increases having negligible effects, suggesting that larger batch sizes reduce wall-clock time in parallel computing environments but increase total simulation effort, emphasizing the need for resource balancing. When ASSR and/or stratification are applicable, we recommend incorporating these features into adaptive sampling methods. Additionally, batch size should be optimized to balance sampling efficiency and computational constraints. Implementing the three features may reduce costs and enable resource reallocation to other traffic safety initiatives, ultimately improving the overall efficiency of virtual safety impact assessment methodologies.

\section*{Acknowledgment}
This work is funded by the SHAPE-IT project under the European Union’s Horizon 2020 research and innovation programme (under the Marie Skłodowska-Curie grant agreement 860410). We would like to thank the European Commission for funding the work. We also extend our gratitude to Volvo Car Corporation for allowing us to use their data and simulation tool, and in particular Malin Svärd and Simon Lundell at Volvo Cars for support in the simulation setup, and Mattias Robertson for his administrative support. Furthermore, we thank Marina Axelson-Fisk and Johan Jonasson at the Department of Mathematical Sciences, Chalmers University of Technology and University of Gothenburg, for their valuable discussions on the methodological aspects of this work.

\bibliographystyle{apalike}

\end{document}